\documentclass[reprint,superscriptaddress,aps,prapplied]{revtex4-1}
\usepackage{amsmath,epsfig,amssymb,subfigure,bm,dsfont}
\usepackage{lipsum} 
\usepackage{color}
\usepackage{graphicx}
\usepackage{dcolumn}
\usepackage{bm}
\usepackage{bbold}
\usepackage{upgreek}
\usepackage{amsfonts,color}
\usepackage{amsmath}
\usepackage[unicode=true, breaklinks=false, pdfborder={0 0 1}, backref=false, colorlinks=true, linkcolor=blue, urlcolor=blue, citecolor=blue]{hyperref}

\newcounter{lastnote}

\begin{document}
\title{Ultrathin bismuth-yttrium iron garnet films with tunable magnetic anisotropy}

\author{Hanchen Wang}
\email{hanchen.wang@mat.ethz.ch}
\thanks{These authors contributed equally to this work}
\affiliation{%
Department of Materials, ETH Zurich, Zurich 8093, Switzerland
}%
\author{William Legrand}
\email{william.legrand@neel.cnrs.fr}
\thanks{These authors contributed equally to this work}
\affiliation{%
Department of Materials, ETH Zurich, Zurich 8093, Switzerland
}%
\affiliation{%
Universit\'e Grenoble Alpes, CNRS, Institut N\'eel, Grenoble 38042, France
}%
\author{Davit Petrosyan}
\affiliation{%
Department of Materials, ETH Zurich, Zurich 8093, Switzerland
}%
\author{Min-Gu Kang}
\affiliation{%
Department of Materials, ETH Zurich, Zurich 8093, Switzerland
}%
\author{Emir Karad\v{z}a}
\affiliation{%
Department of Materials, ETH Zurich, Zurich 8093, Switzerland
}%
\author{Hiroki~Matsumoto}
\affiliation{%
Department of Materials, ETH Zurich, Zurich 8093, Switzerland
}%
\affiliation{%
Institute for Chemical Research, Kyoto University, 6110011 Uji, Japan
}%
\author{Richard Schlitz}
\affiliation{%
Department of Materials, ETH Zurich, Zurich 8093, Switzerland
}%
\affiliation{%
Department of Physics, University of Konstanz, 78457 Konstanz, Germany
}%
\author{Michaela Lammel}
\affiliation{%
Department of Physics, University of Konstanz, 78457 Konstanz, Germany
}%
\author{Myriam H. Aguirre}
\affiliation{%
Instituto de Nanociencia y Materiales de Aragón, CSIC, E-50018 Zaragoza, Spain
}%
\affiliation{%
Dpto. de Física de la Materia Condensada, Universidad de Zaragoza, Pedro Cerbuna 12, E-50009
Zaragoza, Spain
}%
\affiliation{%
Laboratorio de Microscopías Avanzadas, Universidad de Zaragoza, E-50018 Zaragoza, Spain
}%
\author{Pietro Gambardella}
\email{pietro.gambardella@mat.ethz.ch}
\affiliation{%
Department of Materials, ETH Zurich, Zurich 8093, Switzerland
}%
\date{\today}

\begin{abstract} 
We report on the epitaxial growth of nm-thick films of bismuth-substituted yttrium iron garnet (BiYIG) by high-temperature off-axis radio-frequency magnetron sputtering. We demonstrate accurate control of the magnetic properties by tuning of the sputtering parameters and epitaxial strain on various (111)-oriented garnet substrates. BiYIG films with up to -0.80\% lattice mismatch with the substrate remain fully strained up to 60~nm-thick, maintaining a high crystalline quality. Transmission electron microscopy and energy-dispersive X-ray spectroscopy confirm coherent epitaxial growth, the absence of defects, and limited interdiffusion at the BiYIG/substrate interface. Varying the tensile or compressive strain between -0.80\% and +0.56\% in BiYIG allows for accurate compensation of the total magnetic anisotropy through magneto-elastic coupling. The effective magnetic anisotropy of sputtered BiYIG films can be further tuned via the off-axis deposition angle and the oxygen flow during growth, which determine the cation stoichiometry. Under optimized growth conditions, a ferromagnetic resonance (FMR) linewidth of 1~mT at 10~GHz is reliably obtained even for thicknesses as low as 10~nm. We also report small FMR linewidths in ultrathin (2-5~nm) BiYIG films grown on diamagnetic substrate yttrium scandium gallium garnet. 
These findings highlight the promise of low-damping, strain-engineered nm-thick BiYIG films for implementing advanced functionalities in spin-orbitronic and magnonic devices. 
Specifically, the magnetic-anisotropy compensation and low damping enable large cone-angle magnetization dynamics immune to magnon-magnon nonlinear scattering.

\end{abstract}

\maketitle

\section{INTRODUCTION}

Thin films of magnetic garnets are of great importance for applications in spintronics and magnonics as well as for down-scaling of microwave devices~\cite{han2023coherent,zhang2022oxide,csaba2017perspectives,chumak2015magnon,schmidt2020ultrathin}.
Their reduced thickness and low dissipation of magnetic dynamics are key to device performance, especially in schemes employing interfacial spin currents. With its record-low magnetic damping and excellent spin-wave propagation abilities, yttrium iron garnet (YIG) has long stood at the forefront of integrated magnonics~\cite{schmidt2020ultrathin,chumak2015magnon,csaba2017perspectives,demidov2017magnetization,Pirro2021,Kruglyak2010,serga2010yig}. Over the past decade, significant advances in the synthesis of ultrathin YIG films have enabled their integration into chip-scale devices~\cite{yu2014ultralow,chang2014nanometer,wang2014strain,yang2018fmr,fu2017epitaxial,guo2019spin,tang2016exquisite}. Further progress in this area depends on overcoming current material limitations and expanding the functionalities of ultrathin magnetic insulator films~\cite{lammel2022atomic,scheffler2023aluminium,kaczmarek2024atomic,khurana2024rare}. Among several deposition techniques, physical vapor deposition (PVD) based on magnetron sputtering offers scalability and compatibility with industrial processes. However, maintaining precise control of the stoichiometry in ultrathin films remains a key challenge. This aspect is critical for device performance, because the coherence time of the magnetization dynamics is closely related to the cationic composition of the magnetic system. Achieving long-lived magnetization dynamics therefore requires accurate control of the film stoichiometry. Likewise, it requires the mitigation of interfacial intermixing and surface atomic depletion during epitaxial growth, both of which have an increasing impact as magnetic layers get thinner.



Bismuth-substituted YIG (BiYIG) is emerging as a promising material for spintronic and magnonic integrated systems, since it offers combined advantages over unsubstituted YIG~\cite{soumah2018ultra,fakhrul2023substrate,lee2023large,kikkawa2022composition,caretta2020relativistic,fan2023coherent}. The introduction of Bi enhances magneto-optical effects and spin-orbit coupling~\cite{fakhrul2019magneto,wang2025BiYIG,rogalev2009element,li2021magneto,bi2011onchip}, providing access to a wider range of optical and spintronics-based functionalities. BiYIG also enhances the tunability of magnetic anisotropy compared to unsubstituted YIG, via the combination of strain engineering and growth-induced anisotropy mechanisms. This enables precise shaping of the magnetic energy landscape in devices~\cite{song2023engineering}. The possibility of compensation of the effective magnetic anisotropy, or a cancellation of its temperature dependence~\cite{gouere2022temperature}, for example, are highly desirable for spin-orbit torque devices~\cite{merbouche2024true,divinskiy2021evidence,evelt2018emission}. However, BiYIG is a quaternary compound, which complicates the epitaxial growth and its stoichiometry control, even more in PVD conditions. Previous investigations of ultrathin BiYIG epitaxial systems have highlighted their great potential together with persistent challenges in achieving consistent structural and magnetic quality~\cite{soumah2018ultra,fakhrul2023substrate}. Variations in stoichiometry, interface sharpness, and strain have often led to limited dynamic performance, particularly in ultrathin ($<100$~nm) layers, where satisfactory in-plane and out-of-plane ferromagnetic resonance (FMR) properties remain difficult to obtain simultaneously. Therefore, achieving the full potential of BiYIG for integrated magnonics requires a careful understanding of its PVD growth and its optimization, in order to attain a consistent structural quality and magnetic performance.



\begin{figure*}
\hspace{-0.5cm}\includegraphics[width=180mm]{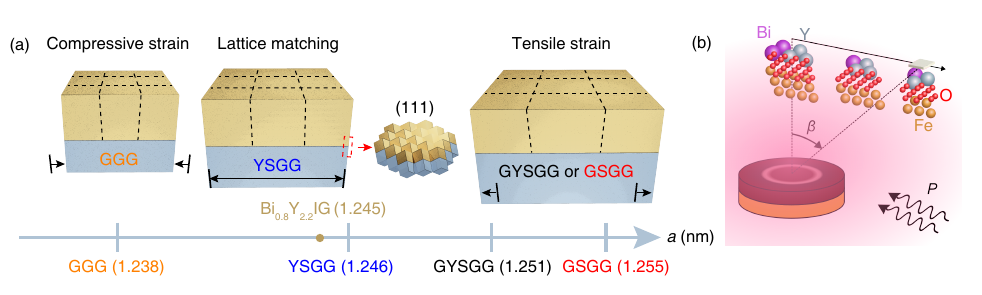}
\caption{(a) Schematic diagrams of the epitaxial relationship of BiYIG with different garnet substrates with (111) orientation. For the nominal composition Bi$_{\rm 0.8}$Y$_{\rm 2.2}$Fe$_{\rm 5}$O$_{\rm 12}$, the GGG substrates provide compressive strain (+0.56\% mismatch with substrate), YSGG is about lattice-matched (-0.08\%), while the GYSGG (-0.48\%) and GSGG (-0.80\%) substrates provide a marked biaxial tensile strain. (b) Schematic representation of the diffusion of atomic species in the deposition plasma for off-axis magnetron sputtering. The angle $\beta$ is the geometric angle between the target center-substrate line and the normal to the target surface. The deposition rate reduces with increasing off-axis angle $\beta$, while the film composition also evolves with $\beta$. Other control parameters are the plasma rf power $P$ and the Ar/O$_{2}$ gas mixture composition.}
\label{fig1}
\end{figure*}



In this work, we investigate the growth of BiYIG thin films using high-temperature off-axis radio-frequency magnetron sputtering, a technique associated to reproducible and scalable fabrication, compatible with industrial processing. Our approach emphasizes precise control over the film composition by tuning growth parameters such as off-axis geometry and partial pressure of oxygen. We employ high-resolution X-ray diffraction, reciprocal space mapping, and transmission electron microscopy to assess epitaxial quality, strain state, and interface sharpness. FMR measurements provide clear insight into how strain and growth conditions influence the key magnetic parameters, such as the effective magnetic anisotropy and Gilbert damping. By growing several (111)-oriented garnet substrates with different lattice parameters, we introduce controlled strain into the films, enabling reproducible and almost continuous tuning of the magnetic anisotropy by strain engineering, while maintaining a very low magnetic resonance linewidth, even in the ultrathin limit down to 2~nm. These results highlight the potential of BiYIG as a reliable and versatile material for spin-orbitronic and magnonic devices. The precise control over its magnetic behavior and its enhanced magneto-optical response are important benefits, while magnetron sputtering offers a practical and reproducible route toward its integration in devices.

\section{METHODS} 
The films are prepared by high-temperature off-axis radio-frequency (rf) magnetron sputtering of a single target with stoichiometric garnet composition  Bi$_{0.8}$Y$_{2.2}$Fe$_5$O$_{12}$ and bulk lattice parameter of about 1.245~nm. Accordingly, we selected four commercial garnet substrates with (111) orientation, so as to investigate growth under lattice-matched, compressive strain, and tensile strain conditions. As shown in Fig.~\ref{fig1}(a), we use Gd$_3$Ga$_5$O$_{12}$ (GGG), Y$_3$Sc$_2$Ga$_3$O$_{12}$ (YSGG), Gd$_{0.63}$Y$_{2.37}$Sc$_2$Ga$_3$O$_{12}$ (GYSGG), and Gd$_3$Sc$_2$Ga$_3$O$_{12}$ (GSGG), with lattice parameters of 1.238~nm, 1.246~nm, 1.251~nm, and 1.255~nm, respectively [see more details in Supplemental Material (SM)~\cite{SM} and Ref.~\cite{legrand2025lattice}]. The lattice mismatch ratio is calculated as $(a_{\rm film} - a_{\rm sub})/a_{\rm sub}$, where $a_{\rm film}$ and $a_{\rm sub}$ are the lattice parameters of the film and substrate, respectively. For each growth, these four substrates are loaded into the magnetron sputtering chamber with a base pressure better than 2$\times$10$^{-5}$ Pa. The sample holder temperature was first increased to 750~$^\circ$C within 15 minutes under an O$_2$ flow of 50~sccm, and maintained for 1 hour at 750 $^\circ$C under 50~sccm O$_2$ for desorption and thermalization, before igniting the target. After pre-sputtering, the sample holder was moved to a position that is off-axis relative to the ignited target, for a deposition with a power of 80~W in a mixed atmosphere of Ar and O$_2$, with respective flow rates of 100 and 5~sccm, and a total pressure of 0.5~Pa, at a sample holder temperature of 750 $^\circ$C. The off-axis deposition method is preferred as it minimizes substrate damage~\cite{yang2018fmr}, and enables fine control of the sample stoichiometry by varying the sample position within the deposition plasma~\cite{yang2018fmr}. As shown in Fig.~\ref{fig1}(b), Fe atoms tend to be deposited in excess, and Bi or Y to be missing at the on-axis position, which inverts at large off-axis angles $\beta$, due to anisotropic diffusion effects in the argon-oxygen plasma, see X-ray photoelectron spectroscopy measurement in SM~\cite{SM,rosenberg2018magnetism,ortiz2018systematic,ortiz2021first}. As a result, the suitable stoichiometry for BiYIG can be achieved by adopting a specific off-axis angle. Such anisotropic diffusion also offers enough tunability in composition if desired, which will be discussed later. After deposition, the films are further annealed in-situ at 750~$^\circ$C under an O$_2$ flow of 50~sccm for 1 hour, and then let to cool down to room temperature under an O$_2$ flow of 50~sccm for an additional hour.

\section{RESULTS AND DISCUSSION} 

\subsection{Structural characterization}

\begin{figure*}
\hspace{-0.5cm}\includegraphics[width=170mm]{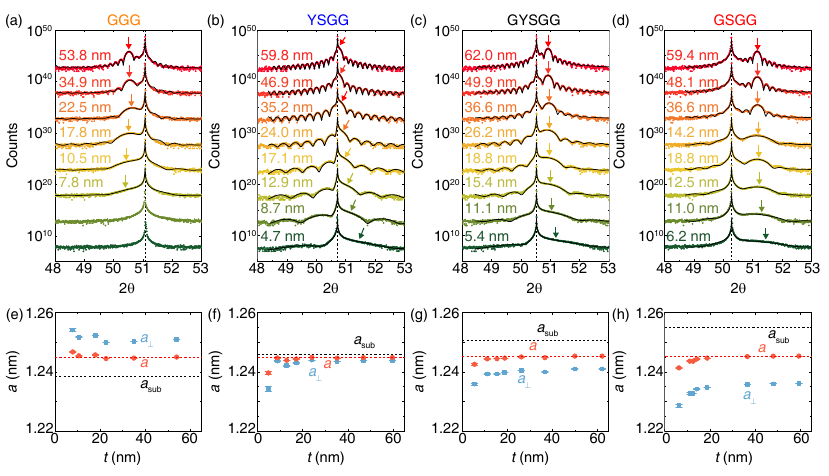}
\caption{(a)-(d) High-resolution XRD with different thicknesses, grown on GGG, YSGG, GYSGG, and GSGG substrates, respectively (each curve is offset by $10^5$ counts). Black curves are the diffractograms fitted to the data using an enhanced kinematical diffraction model. The thickness values are obtained from the Laue-fringe fitting of the high-resolution XRD data. (e)-(h) Thickness dependence of the out-of-plane lattice parameter ($a_{\perp}$) and the equivalent unstrained lattice parameter ($a$), extracted from the fits and compared to the substrate lattice parameter ($a_{\rm{sub}}$).}
\label{fig2}
\end{figure*}

We first investigate the crystalline quality of BiYIG films grown on GGG, YSGG, GYSGG, and GSGG, at an optimal off-axis angle $\beta$ of 39$^\circ$. Figures~\ref{fig2}(a)-(d) show the systematic characterization by X-ray diffraction (XRD) of a thickness series of samples grown on the four substrates (measurements details in SM~\cite{SM}). In these symmetrical 2$\theta-\omega$ diffractograms, the main signal corresponds to the (444) diffraction peak of the substrates, with the $2\theta$ position reflecting their lattice parameters. The (444) peaks from the epitaxial BiYIG films are found near these substrate peaks, and are accompanied by distinct Laue oscillations, which correspond to thickness fringes that indicate a high crystalline quality and a good lateral uniformity in the films. 

To extract the distance $d_{111}$ between diffracting planes and thickness for each sample, the data within $\pm$2 degrees of the 2$\theta$ angle around the film peak are fit with a diffraction model in a kinematic approximation (more details in SM~\cite{SM} and Refs.~\cite{pesquera2011xray,legrand2025lattice}). The black solid curves in Figs.~\ref{fig2}(a)-(d) are fits based on this model, which show excellent agreement with the experimental data for films grown on YSGG and GYSGG. In contrast, for films grown on GGG and GSGG, the central peak is often slightly broadened, and the oscillations decay rapidly with angle. This behavior may result from increased surface roughness or minor inhomogeneities along the thickness direction, leading to the faster loss of Laue fringes with varying angle.

\begin{figure*}
\hspace{-0.5cm}\includegraphics[width=170mm]{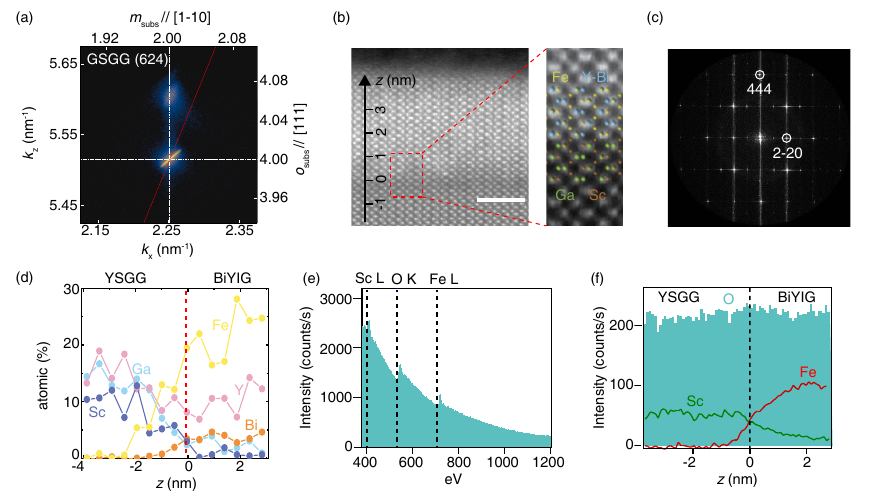}
\caption{(a) X-ray reciprocal space map acquired near the (624) peak for a 60~nm-thick BiYIG film on a GSGG substrate. The vertical and horizontal white dashed lines indicate alignment with the substrate (624) peak within the film plane and along the film normal direction. The red dashed line indicates the peak positions expected in case of a fully relaxed growth. (b) High-angle annular dark-field STEM image of an ultrathin (6~nm) BiYIG film on a YSGG substrate. The scale bar is 2~nm. (c) Fast Fourier Transform (FFT) of the HAADF-TEM image with labeled $444$ and $2\bar{2}0$ peaks. (d) Distribution of each element, obtained by spatially resolved EDX across the YSGG/BiYIG interface [see the detailed position in (b)]. (e) Raw data for EELS and (f) deduced distributions of Sc, Fe and O as a function of position along the film's normal. The YSGG/BiYIG interface is located at $z=0$~nm [see the detailed position in (b)].}
\label{fig3}
\end{figure*}

To verify the growth mode of these films, a reciprocal space map (RSM) was acquired for the 60-nm-thick BiYIG film (the maximum thickness) deposited on GSGG, serving as a representative sample for the entire series, since this sample features the strongest lattice mismatch and the largest thickness. The RSM shown in Fig.~\ref{fig3}(a) displays (624) diffraction peaks from both the substrate and the film, which have a matching in-plane component of the reciprocal vector. This confirms that the BiYIG films are fully strained, showing no evidence of structural relaxation, even for the strongest lattice mismatch and strain investigated here. As a result, these epitaxial films feature rhombohedral distortion, where the unit cell varies in length along the body diagonal aligned with the [111] axis. This extended strain stability up to 60~nm may be attributed to the good stoichiometry control, which promotes plane-by-plane growth and kinetically hinders defects and dislocation formation, allowing the strain to persist to the top of the layers.


The out-of-plane lattice parameter $a_{\perp}$ of the rhombohedrally distorted films, defined as the cubic lattice parameter that would result in the same $d_{111}$ spacing between (111)-oriented planes, is calculated as $a_{\perp}=\sqrt{3}d_{111}$. Owing to the fully strained growth, we can deduce the cubic lattice parameter $a$ expected for films free of strain according to
\begin{equation}
a=a_{\text {subs }}-\frac{\left(1-\mu_{111}\right)}{\left(1+\mu_{111}\right)} \Delta a_{\perp},
\end{equation}
where $a_{\text {subs }}$ is the lattice parameter of the substrate, and $\mu_{111}$ represents the Poisson ratio for (111) oriented films (unknown to us for BiYIG, so we use $\mu_{111}=0.29$ from YIG~\cite{hansen1983magnetic}). The extracted parameters $a_{\perp}$ and $a$ for the different substrates are summarized in Figs.~\ref{fig2}(e)-(h), where the film thicknesses are also directly estimated from the fits. For thicknesses above $\approx$~20~nm, the unstrained lattice parameter $a$ converges on all substrates to 1.245~nm, consistent with the bulk value of Bi$_{0.8}$Y$_{2.2}$Fe$_5$O$_{12}$ (1.245~nm as well). This result reflects the fully strained growth free of structural relaxation. For thicknesses below $\approx$~20~nm, the extracted lattice parameters on all substrates exhibit minor deviations, which we attribute to stoichiometry changes at the interface with the substrate, and at the top surface of the BiYIG layers. 
Such deviations are minor and decrease rapidly with increasing thickness following an approximate $1/t$ trend. However, the amplitude of the lattice parameter change, as well as thickness range over which it occurs, are very different from previous reports of strain relaxation in  YIG~\cite{wang2014strain} or rare-earth iron garnet~\cite{ortiz2018systematic} thin films.

To corroborate these results, scanning transmission electron microscopy (STEM) was performed on a BiYIG film with a nominal thickness of 6~nm, grown on YSGG. A high-angle annular dark-field (HAADF) image is shown in Fig.~\ref{fig3}(b), of which the inset provides a close-up view at the BiYIG/YSGG interface. The film thickness is determined to be 5.6~$\pm$~0.6~nm, corresponding to only 4-5 unit cells of the BiYIG lattice. The Fourier transform analysis in Fig.~\ref{fig3}(c) confirms that the film and substrate share the same crystallographic orientation. No structural defects, dislocations, or stacking faults are observed at the interface, indicating a high-quality epitaxy. A darker contrast is observed at the BiYIG/YSGG interface compared to the substrate, which we attribute to inter-diffusion between the film and substrate, a common characteristic of garnet ultrathin films grown by PVD. Additionally, the film surface appears darker than its center, potentially pointing at variations of stoichiometry at the top surface, to be related to the changes in lattice parameter $a$ exhibited above for the thinner films. 

To investigate this aspect in more detail, energy-dispersive X-ray spectroscopy (EDX) and electron energy loss spectroscopy (EELS) were employed to analyze the concentrations for the different elements, along a scanning line across the interface displayed in Fig.~\ref{fig3}(b). Focusing on the substrate/film interface region, the thickness profiles for Bi, Y, Fe, and substrate elements concentrations obtained by EDX are presented in Fig.~\ref{fig3}(d). The EELS scan only concerns Fe, Sc and O elements, as shown in Fig.~\ref{fig3}(e), and provides a EELS intensity which is increasing with concentration, but not linearly related. The corresponding thickness profiles in Fig.~\ref{fig3}(f) extend from within the YSGG substrate to beyond the top surface of the BiYIG film. Although a precise stoichiometry analysis is difficult for such thin films, a few robust observations are possible. The EDX data reveals a nearly constant atomic concentration of Bi after the interface and a minimal interdiffusion, which constitutes an excellent result considering its volatility. From the EELS data, it is shown that the BiYIG layer has the correct oxygen stoichiometry within the instrumental sensitivity. The inter-diffusion of Sc and Fe is evidenced on a length scale of 1.5--2~nm, consistent with that of the darker contrast in the images, with Sc penetrating more into the film than Fe into the substrate. At the surface, the Fe content decreases within the last 2~nm, and the line scans reveal lower total concentrations, compatible with a roughness of 0.5--1~nm, which accounts for the less bright contrast at the surface compared to the middle of the film in Fig.~\ref{fig3}(b). These observations point at the critical importance of controlling interfacial inter-diffusion and surface composition in ultrathin layers of iron garnet, when addressing their magnetization dynamics.

\subsection{Magnetic anisotropy}

We now move to the investigation of the magnetic anisotropy in sputtered epitaxial BiYIG, focusing first on its dependence on layer thickness and substrate. A custom-built FMR setup was used to monitor the field-derivative microwave transmission signals ($\operatorname{Re} \partial S_{21} / \partial B$ and $\operatorname{Im} \partial S_{21} / \partial B$) through a coplanar waveguide (CPW) holding the films~\cite{Legrand2025}. These field-derivative signals, described by analytical lineshapes deduced from the dynamical magnetic susceptibility, were fitted to extract the dispersion with field of the FMR and thus estimate the magnetic properties of the films, in particular effective anisotropy (more details in SM~\cite{SM}, and Ref.~\cite{Legrand2025}). The linewidth and the dynamical magnetic properties of the films are presented in the next subsection.

\begin{figure}
\hspace{-0.5cm}\includegraphics[width=85mm]{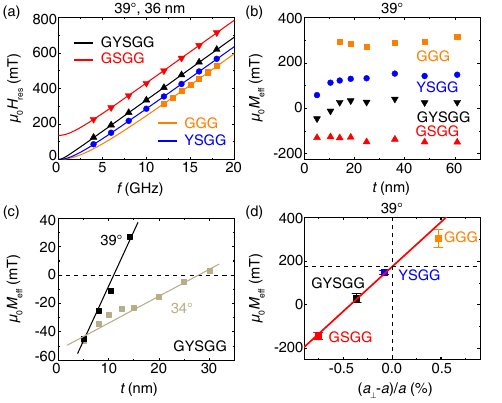}
\caption{(a) Ferromagnetic resonance field $\mu_0 H_{\rm res}$ as a function of the excitation frequency $f$, measured on the 36-nm thick BiYIG films grown on different substrates. The solid curves are fits based on the Kittel formula. (b) Thickness dependence of the effective magnetization $\mu_0 M_{\rm eff}$ estimated from the FMR of the BiYIG films grown on different substrates. (c) Thickness dependence of $\mu_0 M_{\rm eff}$ of BiYIG grown on GYSGG for different off-axis deposition angles $\beta$. (d) $\mu_0 M_{\rm eff}$ in the large thickness limit, as a function of the epitaxial strain. The red line is a linear fit to the data for the four substrates.}
\label{fig4}
\end{figure}

As illustrated in Fig.~\ref{fig4}(a), the resonance fields are obtained as a function of excitation frequency for all series of films grown on the four different substrates. These measurements allow for determining the effective magnetic anisotropy field $\mu_0H_{\text {ani}}$, which consists of a magneto-crystalline component with easy axes along all directions equivalent to [111], and a magneto-elastic component induced by strain along the [111] direction. The analysis employs a Kittel formula neglecting the small magneto-crystalline component, $\omega=\gamma \mu_0 \sqrt{H_{\mathrm{ext}}\left(H_{\mathrm{ext}}+M_{\mathrm{eff}}\right)}$, where $\omega/2\pi$ is the excitation frequency, $\gamma$ is the gyromagnetic ratio, $\mu_0$ is the permeability of free space, $\mu_0 H_{\text{ext}}$ is the external field, and $\mu_0M_{\text {eff }}=\mu_0M_{\text s}-\mu_0H_{\text {ani}}$ is the effective magnetization. The saturation magnetization measurements ($\mu_0M_{\text s}$) for each substrate at large thickness are provided in SM~\cite{SM} to eliminate the possible influence of $\mu_0M_{\text s}$ on $\mu_0M_{\text {eff}}$ for different substrates. As summarized in Fig.~\ref{fig4}(b), the thickness-dependent effective magnetization values $\mu_0M_{\text{eff}}(t)$ are estimated for all films shown above, deposited at an off-axis angle of 39$^\circ$. Notably, $M_{\text{eff}}$ deviates from a base line when the film thickness reaches below 20~nm. For films grown on YSGG and GYSGG, the trends of $M_{\text {eff}}$ with decreasing thickness align well with the lattice parameters derived from XRD measurements: a larger negative lattice mismatch reinforces strain-induced anisotropy and reduces $M_{\text{eff}}$. In addition, $M_{\text s}$ decreases at low $t$, causing $M_{\text{eff}}$ to reduce even further (see more details in SM~\cite{SM}). In contrast, for films grown on GSGG and GGG, $M_{\text{eff}}(t)$ shows additional features that compensate this trend. 

Below 20~nm, the effective magnetization of the BiYIG films grown on GYSGG gradually crosses zero, indicating a compensation of the effective magnetic anisotropy. To further investigate the regime of compensated magnetic anisotropy, in which the perpendicular magnetic anisotropy balances the shape anisotropy arising from the demagnetizing field the thin-film ($M_{\text{eff}}=0$), additional samples with thicknesses between 5 and 30~nm were grown on GYSGG again, motivated by the observation in Fig.~\ref{fig4}(b) that the $\mu_0M_{\text {eff}}$ of BiYIG grown on this substrate crosses zero slightly above 10~nm. Their effective magnetization values are summarized in Fig.~\ref{fig4}(c). At an off-axis angle of 39$^\circ$, the anisotropy compensation occurs around $t=$~11~nm. By reducing the off-axis angle to 34$^\circ$, the evolution of $\mu_0M_{\text {eff}}$ with thickness is slower, and the compensation thickness shifts close to 28~nm. This shift is essentially related to changes in the stoichiometry of the BiYIG films grown at different positions within the plasma. As shown in Fig.~\ref{fig1}(b), a smaller off-axis angle leads to a more pronounced accumulation of Fe relative to Y and Bi (see also SM~\cite{SM}). Thereby, this affects the lattice parameter and the saturation magnetization of the BiYIG films deposited closer to the on-axis position. The change in stoichiometry leads to a slower saturation of the out-of-plane lattice parameter with increasing film thickness. In addition to the off-axis angle, the oxygen flow during deposition provides an additional degree of control over the magnetic anisotropy. This aspect is discussed more extensively in SM~\cite{SM}.

Next, we examine further the large thicknesses regime, characterized by a nearly constant effective magnetization and constant out-of-plane lattice parameter. To quantify the strain, we use the vertical strain ratio $a_{\perp}/a-1$. As shown in Fig.~\ref{fig4}(d), the effective magnetization and the vertical strain ratio are linearly related for samples with $t>$~20~nm. The line represents a linear fit to the data on the four substrates, with an intercept of +175 $\pm$ 13~mT, consistent with the saturation magnetization $\mu_0M_{\rm s}$ of BiYIG. 
This corresponds to a change of $\mu_0M_{\rm eff}$ of 410~$\pm$~30~mT per 1\% of strain. This sign of the slope is consistent with the magneto-elasticity in (Bi,Y)IG, and the large range of tunability obtained here highlights the flexibility of BiYIG to adapt to various requirements.
Notably, Fig.~\ref{fig4}(d) reflects the dominant first-order strain-induced anisotropy in BiYIG, manifested by the linear relationship between the vertical strain ratio and 
$\mu_{0}M_{\mathrm{eff}}$. We emphasize, however, that garnet materials may also host higher-order strain-driven anisotropy terms, as discussed for EuIG in Ref.~\cite{ortiz2021first}. Although such contributions cannot be resolved within the present in-plane FMR geometry, their investigation represents a valuable direction for future work.

\subsection{Magnetization dynamics}

\begin{figure}[t]
\hspace{-0.5cm}\includegraphics[width=85mm]{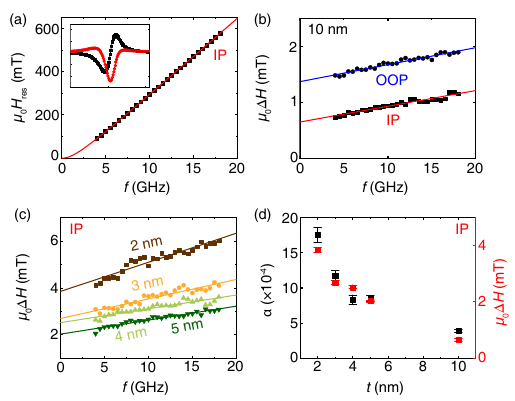}
\caption{(a) In-plane FMR field $\mu_0 H_{\rm res}$ as a function of excitation frequency measured for a 10-nm thick BiYIG film deposited on YSGG with an off-axis angle of 39$^\circ$. The inset shows the measured derivatives of the complex microwave transmission $\operatorname{Re},\operatorname{Im} \partial S_{21} / \partial B$ as a function of external magnetic field around resonance. (b) Resonance linewidth $\mu_0 \Delta H$ as a function of frequency, measured for either in-plane and out-of-plane FMR. The solid lines are fits to Eq.~\eqref{eq:LW}. (c) Resonance linewidth $\mu_0 \Delta H$ as a function of frequency measured for BiYIG films on YSGG with thickness 2~nm, 3~nm, 4~nm, 5~nm, in-plane FMR. (d) Thickness dependence of Gilbert damping $\alpha$ and inhomogeneous linewidth $\mu_0 \Delta H_0$ of BiYIG on YSGG.}
\label{fig5}
\end{figure}

Finally, we demonstrate that the BiYIG epitaxial films exhibit excellent microwave-frequency dynamical properties even when their thickness is less than 10~nm, making them highly suitable for magnonics. We report in Figs.~\ref{fig5}(a) and (b) the FMR measurement of a 10-nm-thick BiYIG film grown on YSGG with an off-axis angle of 39$^\circ$. The spectra display the derivative Lorentzian line shapes expected from a single resonance peak, see inset of Fig.~\ref{fig5}(a), thus enabling reliable fits of the resonance linewidth. This lineshape further confirms the very satisfactory film homogeneity. The gyromagnetic ratio $\gamma/2\pi=$~28.134~$\pm$~0.005 GHz/T is obtained from Fig.~\ref{fig5}(a) by a fit to the Kittel formula. The FMR linewidth as a function of frequency, under either in-plane (IP) or out-of-plane (OOP) external magnetic fields, are summarized in Fig.~\ref{fig5}(b), and fitted to
\begin{equation}
\label{eq:LW}
\mu_0 \Delta H=\mu_0 \Delta H_0+2 \alpha \omega / \gamma.
\end{equation}
This gives an inhomogeneous broadening contribution to the linewidth $\mu_0 \Delta H_0 = 0.66 \pm 0.05$~mT for IP FMR and $1.37 \pm 0.02$~mT for OOP FMR. The corresponding Gilbert damping parameters are $\alpha = 0.00039 \pm 0.00002$ for IP FMR and $\alpha = 0.00041 \pm 0.00002$ for OOP FMR. In addition to this 10~nm-thick BiYIG film, we also measured the damping and inhomogeneous broadening for ultrathin BiYIG films on YSGG with thicknesses ranging from 2 to 5~nm, in 1~nm increments. The IP FMR linewidth in these films is shown as a function of frequency in Fig.~\ref{fig5}(c). Overall, both the inhomogeneous broadening and the Gilbert damping increase as $1/t$, rather than $1/t^2$, with decreasing film thickness $t$, as summarized in Fig.~\ref{fig5}(d). This trend is mainly attributed to enhanced magnon-magnon scattering at ultralow thicknesses~\cite{schmidt2020ultrathin}. We also provide the raw FMR spectra for the 2~nm-thick sample in SM~\cite{SM}. They illustrate that the line-shape remains nearly ideal and well-defined, despite the ultrathin film thickness. The inhomogeneous broadening and the Gilbert damping values reported here belong to the lowest ones in comparison to high-quality, $<10$~nm-thick unsubstituted YIG films~\cite{schmidt2020ultrathin}. 


\section{CONCLUSIONS}

In summary, we have demonstrated the epitaxial growth of high-quality BiYIG thin films on various (111)-oriented garnet substrates with high-temperature off-axis radio-frequency magnetron sputtering. High-resolution XRD, RSM, and STEM investigation confirmed that these films undergo coherent growth and remain fully strained up to at least 60~nm in thickness. They also present excellent crystallinity, homogeneous Bi concentration across the thickness, and well-defined interfaces with minimal defect formation. By selecting substrates with suitable lattice parameters, the introduction of compressive or tensile strain in the films enables systematic control of the magnetic anisotropy, in addition to thickness-dependent effects. FMR measurements revealed that the effective magnetic anisotropy of these BiYIG films can be finely tuned to compensation, through the adjustment of strain, deposition parameters (such as off-axis angle and oxygen flow), and film thickness.

Most importantly, these films combine low damping and very moderate inhomogeneous broadening, comparable to YIG layers of similar thicknesses. Different from YIG, however, BiYIG films allow for magnetic anisotropy tunability through strain engineering, and unlock large signals in magneto-optical investigations (see more details in SM~\cite{SM}). The unique combination of low damping in films with sub-10-nm thickness, widely tunable anisotropy, and strong magneto-optical effects makes them appealing for exploring novel directions in magnonics, associating, for example, spin injection, tailored magnon band structures, and optical readout~\cite{Wang2025magnoncavity}. 
Such tailored growth of ultrathin magnetic insulators with nearly compensated anisotropy is also relevant for the investigation of dynamical behaviors in chiral magnetic textures, for which BiYIG provides an excellent host material~\cite{Wang2025chiral}. Thicker films, offering even lower damping, are highly relevant for optical manipulation of magnetization dynamics~\cite{wiechert2026laser}.

Finally, BiYIG grown on lattice-matched YSGG prevents the diffusion of paramagnetic atoms from the substrate, which has been shown to negatively affect the properties of YIG grown on GGG~\cite{suturin2018gallium,roos2022ggg}. The present investigation not only highlights how to deal with the PVD growth constraints affecting iron garnet films at the ultrathin limit, but also provides a robust framework to engineer BiYIG thin films with tailored functionalities. The combination of tunable anisotropy, excellent magnetization dynamics, and enhanced magneto-optical response positions BiYIG as a highly promising material for next-generation spin-orbitronic and magnonic circuits.

\begin{acknowledgments}

We thank T.~Weber for his assistance in acquiring and interpreting the X-ray diffractograms, and G.~Berthom\'{e} for assistance in acquiring X-ray photoelectron spectroscopy data. This research was supported by the Swiss National Science Foundation (Grant No.~200021-236524). H.W.~acknowledges the support of the China Scholarship Council (CSC, Grant No.~202206020091). W.L.~acknowledges support from the ETH Zurich Postdoctoral Fellowship Program (21-1 FEL-48). H.M.~acknowledges support from JSPS Postdoctoral Fellowship (Grant No.~23KJ1159) and Swiss Government Excellence Scholarships 2024-2025. R.S.~and M.L.~acknowledge funding by the Deutsche Forschungsgemeinschaft (DFG, German Research Foundation) – project number 425217212. M.H.A.~acknowledges the financial support of European Commission through Marie Skłodowska-Curie Actions H2020 RISE with the project ULTIMATE-I (Grant No.~101007825), and the access to equipment of “Servicio General de Apoyo a la Investigación (SAI), Universidad de Zaragoza".
\end{acknowledgments}

\appendix

\section{Optical constant measurement by magneto-optic Kerr effect}

\begin{figure}[htp]
\hspace{-0.5cm}\includegraphics[width=85mm]{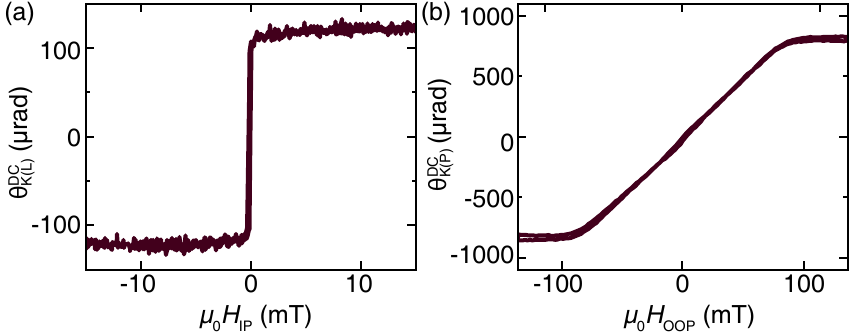}
\caption{In-plane (a) and out-of-plane (b) magnetic hysteresis loops measured by longitudinal and polar MOKE, respectively to characterize the magneto-optical constants of the BiYIG. }
\label{fig9}
\end{figure}

In this section, we performed magneto-optic Kerr effect (MOKE) measurements on a $5~\mathrm{nm}$-thick BiYIG film grown on a YSGG substrate to characterize its magneto-optical constants. A laser with wavelength of $520~\mathrm{nm}$ was employed, illuminating the sample from the top. Both longitudinal and polar MOKE configurations were used to obtain the Kerr rotation angles $\mathrm{\theta_K}$ for in-plane (IP) and out-of-plane (OOP) magnetic hysteresis loops, respectively, aiming to find $\alpha_\mathrm{MO}=\max \mathrm{\theta_K} /(\mu_0 M_{\text s})$, in each case. As shown in Fig.~\ref{fig9}, large Kerr rotation angles are observed in both geometries, with the signal saturating at approximately $120~\mathrm{\upmu rad}$ and $800~\mathrm{\upmu rad}$ for the IP and OOP measurements, respectively. By dividing these values by the measured saturation magnetization in SM~\cite{SM}, we obtain the IP and OOP magneto-optical constants as $\alpha_\mathrm{MO}^\mathrm{IP}\approx 0.69~\mathrm{\upmu rad/mT}$ and $\alpha_\mathrm{MO}^\mathrm{OOP}\approx 4.57~\mathrm{\upmu rad/mT}$, respectively. In contrast, pure YIG exhibits almost no detectable Kerr signal at this wavelength due to its vanishing magneto-optical constant~\cite{hansen1983magnetic, kehlberger2015enhanced}. The pronounced Kerr response in BiYIG highlights its strong magneto-optic properties, related to the enhanced spin–orbit coupling induced by Bi substitution.

\bibliographystyle{unsrt}

\end{document}